\documentclass[fleqn,usenatbib]{mnras}

\usepackage{newtxtext,newtxmath}

\usepackage{graphicx}
\usepackage{booktabs} 
\usepackage{amsmath}
\usepackage{tikz}
\usepackage{multirow}
\usepackage{hyperref}
\newcommand{\Sref}[1]{\hyperref[#1]{Section~\ref*{#1}}}
\newcommand{\eref}[1]{\hyperref[#1]{Eq.~(\ref*{#1})}}

\usepackage[T1]{fontenc}
\usepackage{lineno}
\DeclareRobustCommand{\VAN}[3]{#2}
\let\VANthebibliography\thebibliography
\def\thebibliography{\DeclareRobustCommand{\VAN}[3]{##3}\VANthebibliography}

\title[Cosmic Voids and the Puzzle of GRB 221009A]{Influence of Cosmic Voids on the propagation of TeV Gamma Rays and the Puzzle of GRB 221009A}

\author[H. Abdalla et al.]{
Hassan Abdalla$^{1,2,3}$\thanks{E-mail: hassanahh@gmail.com},
Soebur Razzaque$^{4,5,6}$,
Markus Böttcher$^{3}$,
Justin Finke$^{7}$ and
Alberto Dom\'inguez$^{1}$
\\
$^{1}$ IPARCOS and Department of EMFTEL, Universidad Complutense de Madrid, E-28040 Madrid, Spain\\
$^{2}$ Department of Astronomy and Meteorology, Omdurman Islamic University, Omdurman 382, Sudan\\
$^{3}$ Centre for Space Research, North-West University, Potchefstroom 2520, South Africa \\
$^{4}$ Centre for Astro-Particle Physics (CAPP) and Department of Physics, University of Johannesburg, PO Box 524, Auckland Park 2006, South Africa\\
$^{5}$ Department of Physics, The George Washington University, Washington, DC 20052, USA\\
$^{6}$ National Institute for Theoretical and Computational Sciences (NITheCS), South Africa \\
$^{7}$ U.S.\ Naval Research Laboratory, Code 7653, 4555 Overlook Ave.\ SW, Washington, DC, 20375-5352, USA
}

\date{Accepted XXX. Received YYY; in original form ZZZ}

\pubyear{2024}

\begin{document}
\label{firstpage}
\pagerange{\pageref{firstpage}--\pageref{lastpage}}
\maketitle

\begin{abstract}
The recent detection of gamma-ray burst GRB~221009A has attracted attention due to its record brightness and first-ever detection of $\gtrsim 10$ TeV
$\gamma$ rays from a GRB. Despite being the second-nearest GRB ever detected, at a redshift of $z=0.151$, the distance is large enough for severe attenuation of $\gamma$-ray flux at these energies due to $\gamma\gamma\to e^\pm$ pair production with the extragalactic background light (EBL). 
Here, we investigate whether the presence of cosmic voids along the line of sight can significantly impact the detectability of very-high energy (VHE, $>$ 100 GeV) gamma rays from distant sources. Notably, 
we find that the gamma-gamma opacity for VHE gamma rays can be reduced by approximately 10\% and up to 30\% at around 13 TeV, the highest-energy photon detected from GRB~221009A, for intervening cosmic voids along the line-of-sight with a combined radius of 110 Mpc, typically found from voids catalogs, and 250 Mpc, respectively. This reduction is substantially higher for TeV photons compared to GeV photons, attributable to the broader target photon spectrum that TeV photons interact with. This finding implies that VHE photons are more susceptible to variations in the EBL spectrum, especially in regions dominated by cosmic voids. Our study sheds light on the detection of $\gtrsim 10$ TeV gamma rays from GRB 221009A in particular, and on the detection of extragalactic VHE sources in general.
\end{abstract}




\section{Introduction} \label{sec:intro}
Gamma-rays from sources located at a cosmological distance and with energies greater than the threshold energy for electron-positron ($e^\pm$) pair production can be annihilated due to $\gamma \gamma$ absorption by low-energy extragalactic background photons. These intergalactic $\gamma\gamma$ absorption signatures have attracted great interest in astrophysics and cosmology due to their potential to indirectly measure the Extragalactic Background Light \citep[EBL, e.g.,][] {Finke2009ApJ, SC9, Dominguez11a, Dominguez11c, Sinha2014ApJ, HESS:2017vis, CTA:2020hii}, and thereby probe the cosmic star-formation history \citep[e.g.,][]{Dwek2013APh, Fermi-LAT2018Sci}. Extensive research has been conducted to study the $\gamma\gamma$ absorption imprints on the gamma-ray fluxes from distant sources \citep[e.g.,][]{Aharonian06, Razzaque2009ApJ,  Kneiske2010A&A, Finke10, Gilmore12, Dom15, BW15, Tavecchio16, Dzhatdoev17, Martinez2018hpt, Korochkin2018, Dzhatdoev19, Lang19, Long20, Troitsky2020, LI21, Terzi2021, Finke2022, AZZAM2023, PerezdelosHeros2022izj, Qin2023wot, AlvesBatista:2023wqm}. Observations have revealed that the spectra of some distant blazars at very high energies (VHE; E $>$ 100~GeV) appear to be unexpectedly harder after correcting for $\gamma\gamma$ absorption by the EBL \citep[e.g.,][]{Furniss13,Furniss15, Abdalla17}. One potential explanation for the observed spectral hardening could be a lower density of the EBL than what is currently estimated by the existing models. The EBL might exhibit an inhomogeneous distribution, with the presence of cosmic voids along the line of sight to distant blazars. These voids could feasibly lessen the attenuation of gamma rays emitted by these blazars, thereby rendering them detectable at greater redshifts \citep[e.g.,][]{Furniss13, Abdalla17,Kudoda17, Kudoda:2018pqg}. 

The recent detection of VHE $\gamma$ rays from the gamma-ray burst GRB~221009A at a redshift of $z = 0.151$ \citep{Ugarte-Postigo_2022, Williams2023ApJ, Frederiks:2023bxg,2023ApJ...952L..42L, Burns:2023oxn, Klinger:2023nus,2023SCPMA..6689511W, HESS:2023yeg, Zhao:2022wjg, 2023ApJ...946L..27A, 2022arXiv221013052X, Zheng:2023xsl, 2024arXiv240312851Z, 2024ApJ...966...31K, 2024arXiv240515855B, Ror:2024vgh, 2024A&A...683A..55C}
by the Large High Altitude Air Shower Observatory~\citep[LHAASO, ][] {Huang_2022, LHAASO23} has increased focus on EBL models and fundamental physics, since the $\gamma\gamma$ absorption effect is expected to reduce the source flux by a factor as large as $\sim 2.2\times 10^{4}$ at 18~TeV \citep{Finke2022arXiv221011261F}, the energy of the highest-energy photon detected by LHAASO. 
In its recent publication \cite{LHAASO2023}, the LHAASO Collaboration conservatively estimated the highest-energy photon to be a 13~TeV or 18~TeV photon, depending on whether the intrinsic spectrum is a power-law function with an exponential cutoff or a log-parabola, respectively, after considering the systematic uncertainties in the photon-energy determination.
In order to rigorously evaluate various EBL models, the study detailed in \cite{LHAASO2023} adopts a methodological approach. This involves initially characterizing the intrinsic spectrum of Gamma-Ray Bursts (GRBs) utilizing a log-parabolic (LP) function, $dN/dE =  J_{0} (E/1\rm TeV)^{-a- b\log(E/1\rm TeV)}$, or a power-law function with exponential cutoff (PLEC), $dN/dE =  J_{0} (E/1\rm TeV)^{-\alpha} e^{-E/E_{\rm cut}}$, and subsequently, obtaining the parameters of the functions from fitting LHAASO data. 
This systematic procedure facilitates a comprehensive and quantitative analysis of the different EBL models under consideration. Precise parameters were obtained by \cite{LHAASO2023} from the fitting process for the time intervals 230-300~s and 300-900~s after $T_{90}$.

Although there is a non-zero probability that the highest-energy photon originated from a Galactic HAWC source \citep[e.g.,][]{fraija22}. Nevertheless, explaining even an $\gtrsim 10$~TeV photon from this source is difficult. One possibility is via Ultra-high energy cosmic rays (UHECRs) emitted from GRB~221009A which produce $\gamma$ rays by interacting with CMB and EBL photons along our line-of-sight \citep[e.g.,][]{Das2023A&A...670L..12D, Batista2022arXiv221012855A, 2023ApJ...947...84H, He:2024xqd, Kalashev:2024ayr}. Exotic physics involving Lorentz invariance violation~\citep[LIV, e.g.,][]{dzhappuev22, baktash22,li22, Finke2022arXiv221011261F, Zheng23-83001, 2024CQGra..41a5022A, Yang:2023kjq, LHAASO:2024lub} or mixing of $\gamma$ rays with Axion-like particles~\citep[ALPs, e.g.,][]{baktash22, troitsky22, galanti22, Wang23okw, Gonzalez:2022opy, Nakagawa:2022wwm, Gao23, Troitsky2023uwu, Rojas:2023jdd} or radiative decay during the propagation of a hypothetical heavy neutrino produced by GRB~221009A \citep[e.g.,][]{Cheung2022arXiv221014178C, Smirnov2022arXiv221100634S, Huang:2022udc,2023PhRvD.108b1302G, Brdar:2022rhc, 2024arXiv240405354A, 2024PhRvD.109f3007D, 2023arXiv230814483S} have also been proposed to evade $e^\pm$ pair-production with EBL photons.

In this work, we seek to explain the detection of tens of TeV $\gamma$ rays from GRB~221009A as a result of a reduced EBL intensity, and hence weaker $\gamma\gamma$ absorption, due to presence of a cosmic void, regions devoid of stars and dust, along the line-of-sight. In previous works \cite{Abdalla17, Abdalla18, Abdalla:2018H7}, we investigated the effect of such a cosmic void along the line-of-sight to a distant $\gamma$-ray source on the resulting angle-averaged EBL energy density by considering the inhomogeneity of the direct starlight component of the EBL model by \citet{Razzaque2009ApJ} in the optical-ultraviolet. In the present work, we also consider the inhomogeneity of the dust component of the EBL in the infrared-optical and accordingly modify the model by \citet{Finke10}. We also considered the potential existence of cosmic voids, incorporating \cite{Dominguez11a}, \cite{Saldana-Lopez23} and \cite{Finke2022} EBL models. Specifically, we examined the effect of a reduced $\gamma$-ray opacity due to the presence of such voids on the intrinsic spectra of GRB~221009A. We organize the paper in the following way.
In Section \ref{void_catalog}, we summarize observations of voids from large-scale surveys and perform a Monte Carlo study to estimate the potential amount of void that could exist between us and a source located at the same redshift as GRB 221009A. In Section \ref{Model}, we describe the model setup and the method used to evaluate the EBL characteristics and the resulting $\gamma\gamma$ opacity. The results and conclusions are presented in Sections \ref{res} and \ref{Conclusions}, respectively. 
Throughout this paper the following cosmological parameters are assumed: 
$H_{0} = 70$~km~s$^{-1}$~Mpc$^{-1}$, $\Omega_{m} = 0.3$, $\Omega_{\Lambda} = 0.7$.
\section{The Presence of Cosmic Voids in Large-Scale Cosmic Surveys} \label{void_catalog}
\subsection{Observational considerations}
In recent years, there has been an intensive study of the large-scale structure and dynamics of galaxy distribution. The role of cosmic voids as dynamically distinct elements in the universe's large-scale structure has been firmly established through both theoretical and observational frameworks~\citep[e.g.,][]{Hoffman1982ApJ,  SDSS2008tqn, pan2012m, Tavasoli:2012ai, SUTTER20151, Wojtak2016brz, Kelly2022}. 
The most extensive void catalog currently available is designed to enable precise cosmological experiments using voids and to facilitate more accurate testing of galaxy formation theories is SDSS DR7~\citep[e.g.,][]{SUTTER20151, Kelly2022}. Characterizing cosmic voids provides distinct insights into the large-scale distribution of galaxies, their evolution, and the prevailing cosmological paradigm~\citep{Micheletti2014}. Traditionally, the quest to identify voids within galactic distributions has employed one of two methodologies: firstly, by Identifying empty spherical regions amidst the galactic expanse, or secondly, through the implementation of a watershed algorithm designed to interconnect regions of low density. The preeminent catalogs of cosmic voids, utilized extensively in contemporary astrophysical research, have been derived primarily from the seventh data release of the Sloan Digital Sky Survey SDSS DR7 \citep[see,][]{SDSS2008tqn, SUTTER20151}. Recently,~\cite{Kelly2022} developed a series of cosmic void catalogs, utilizing data from the SDSS DR7. Their approach incorporated three sophisticated void-finding algorithms: VoidFinder and two algorithms based on the ZOBOV framework \citep{Neyrinck:2007gy}, namely VIDE and REVOLVER. This comprehensive analysis led to the identification of over 500 cosmic voids, each exhibiting radii in excess of 10 Mpc/h. The research delved into various characteristics of these voids, including their radial density profiles and the spatial distribution of galaxies within them. Furthermore, this study's results were cross-referenced with simulated data from the Horizon Run 4 N-body simulation's mock galaxy catalogs, demonstrating a remarkable congruence that substantiated the efficacy of their methodologies and the validity of their findings. 
\begin{figure*}
\includegraphics[width=\textwidth, angle=0]{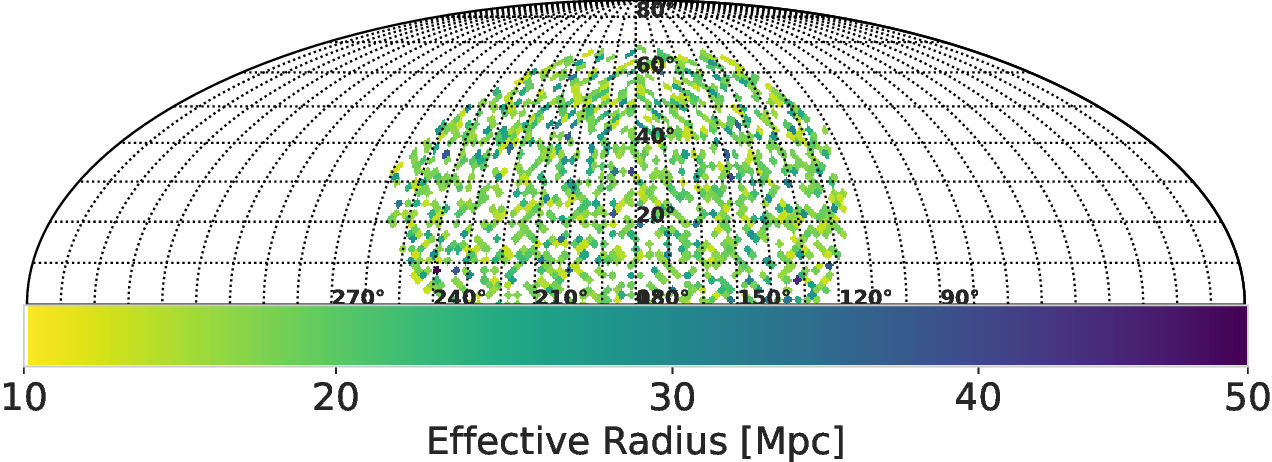} 
 \caption{
This illustration, based on the updated VoidFinder SDSS DR7 from~\citet{Kelly2022}, shows the distribution of 2,303 cosmic voids with effective radii between 14.53 to 43.91 Mpc. The colors on the map reflect the voids' relative comoving effective radii, weighted by radius and plotted using HEALPix grid cells to ensure consistent solid angle coverage across the sky. This method accurately represents spatial distribution by assigning weights to corresponding pixels on the HEALPix map. This is done by using a HEALPix map with mollview for a Mollweide projection in Celestial Equatorial coordinates.}
\label{voids}
\end{figure*} 
\begin{figure*}
\includegraphics[width=\textwidth, angle=0]{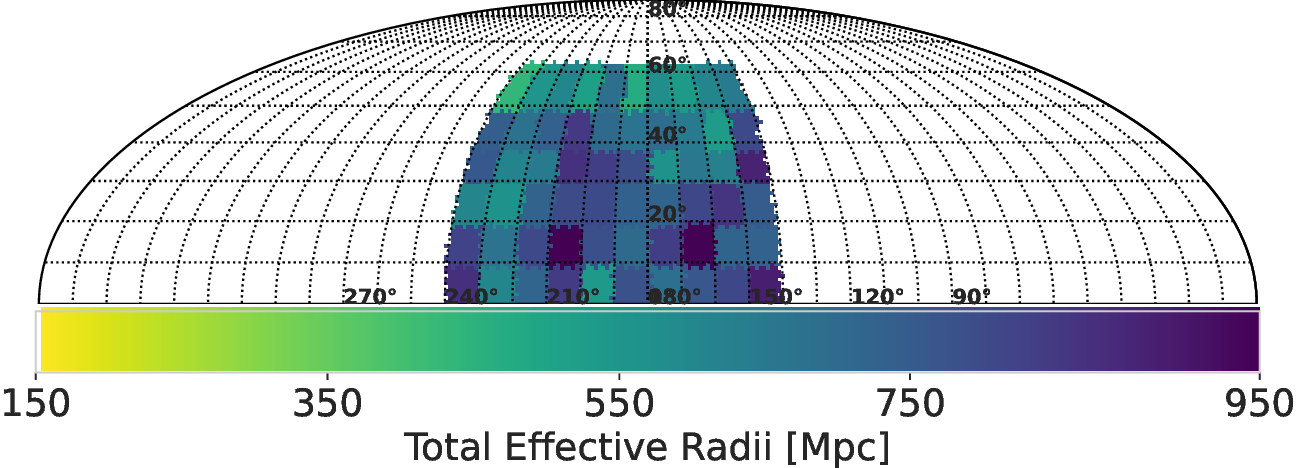}   
 \caption{The density of voids within a specified solid angle on the sky is quantified by calculating the total comoving effective radii within each grid cell shown in Figure~\ref{voids}. Each cell covers a 10-degree square in both Right Ascension and Declination. These cells are weighted based on the cumulative comoving effective radius encompassed within. The color bars represent the relative amount of void content in each cell compared to others, visually illustrating variations in void density across the sky. This is done by using a HEALPix map with mollview for a Mollweide projection in Celestial Equatorial coordinates.}
\label{hist}
\end{figure*} 
To provide a clear visual representation, we have plotted the 
locations of voids of different sizes in the sky in Figure~\ref{voids} that demonstrates the quantity of voids within each grid cell. These cells specifically fall within the ranges of 140$^\circ$ to 240$^\circ$ in Right Ascension (RA) and 0$^\circ$ to 60$^\circ$ in Declination (Dec). Each cell depicted in Figure~\ref{voids}, encompasses the cumulative voids (total effective radius) up to a redshift of $z=0.114$. This value represents the maximum redshift encompassed by the VoidFinder updated Void Catalogs from~\citet{Kelly2022}. It is important to note that the sky region encompassing the extraordinarily luminous GRB 221009A, positioned at RA = 288.264° and Dec = 19.768°, is not included in these catalogs. Consequently, accurately estimating the extent of voids along our line of sight to GRB 221009A is currently unfeasible. {To visualize the variations of cumulative void sizes in different locations of the sky, we present Figure 1, which illustrates the distribution of relative void sizes within a region covered by the updated Void Catalogs SDSS DR7~\citep{Kelly2022} using the VoidFinder algorithm.} 
In Figure~\ref{hist}, we computed the accumulated comoving effective radii contained in the volume of the solid angle subtended by each cell within the celestial coordinates of 140 degrees to 240 degrees in Right Ascension and 0 degrees to 60 degrees in Declination. A notable finding is that there are regions significantly under-dense, with some cells containing around three times more voids than others. Our choice of a 10-degree square grid is arbitrary. At smaller scales, such as within 5-degree square cells, variations become more pronounced. We found that the minimum accumulated effective radii within a grid cell can be as low as zero, while the maximum, for the grid cell containing the most voids, can reach up to approximately 250 Mpc. 
The goal of using a 10-degree grid in Figure~\ref{hist} is to demonstrate that significant variations in void quantities can be found even at such large scales. Therefore, it is conceivable that GRB 221009A might have been situated in a direction where an under-dense region exists between us and the source. Such a condition could result in reduced EBL energy density, potentially facilitating the arrival of TeV photons. 
In Section \ref{Model} we present detailed calculation by assuming an arbitrary possible void sizes, using the \citet{Finke10} EBL model. We characterize the voids in the rest of this Section.
\subsection{Monte Carlo study on the average radii of cosmic voids} \label{Monte Carlo}
The objective of Figures \ref{voids} and~\ref{hist} is to provide a general understanding based on observations that the distribution of cosmic voids varies; certain areas of the sky possess a higher concentration of voids in comparison to others.
Here, we proceed to obtain a more precise quantitative estimate of the cumulative void radius within a spherical region in the direction towards a source, from an observational perspective. This involves assessing the likelihood of encountering such amount of voids in a specific area of the sky where a particular source might be situated behind these voids. To estimate the potential cumulative void radius of voids between us and a source at redshift $0.151$, such as GRB 221009A, we initially extrapolated observed
data beyond $z = 0.114$, the limit covered by the most recent Void Catalogs from the VoidFinder SDSS DR7, extending up to $z = 0.151$.
This was done by assuming that the distribution of voids follows the same pattern up to this redshift. The properties of voids 
are not expected to change significantly over such low redshifts~\citep[see,][]{Sheth:2003py, 2011IJMPS...1...41V, Paz:2013sza}.
The primary goal of this work is to determine if voids exist along the line-of-sight to a very high energy (VHE) gamma-ray source, potentially leading to a reduction in the EBL density and, consequently, the optical depth for TeV photons. In Section \ref{Model}, for simplicity, we assumed the possibility of multiple voids with typical radii distributed in a large spherical region between the Earth and the source. This section explores the feasibility of this assumption from an observational consideration. However, the voids catalogs of SDSS do not contain the portion of sky where GRB{221009A} is located; therefore, we cannot obtain an exact realistic estimation of the total effective radius of voids towards the source. 
{
Therefore, we estimate the likelihood of such an amount of voids being present in a large spherical region between us and the source at $z = 0.151$. To achieve this, we conducted a comprehensive Monte Carlo simulation using the following approach:
}
\begin{figure}
\centering
 \includegraphics[width=\columnwidth, angle=0]{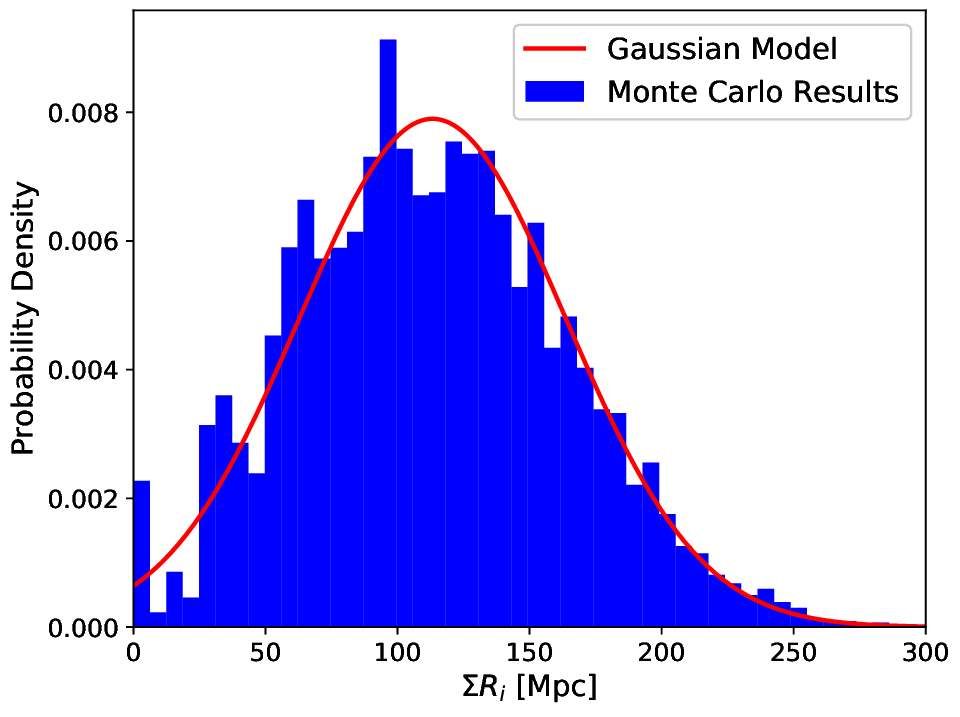}
 \caption{Results of Monte Carlo simulations of the cumulative effective radii of voids within spherical regions, situated between us and various random coordinates. These coordinates fall within the Right Ascension (RA) range of 140 to 230 degrees and the Declination (Dec) range of 10 to 60 degrees. The Gaussian distribution, depicted by the red line, has a mean of 113 Mpc and a standard deviation of 50 Mpc.
\label{Monte_Carlo}}
\end{figure} 
 {We randomly selected points within specified celestial coordinates, ranging from 140 to 230 degrees in RA and 10 to 60 degrees in Dec at redshift z = 0.15. For each iteration, we draw a spherical region with its diameter extending from the observer to one of these points. Within this spherical comoving volume, we calculated the cumulative comoving effective radius ($\Sigma R_{i}$), which is found in the VoidFinder {SDSS DR7 void catalog} and our extrapolated data beyond redshift z = 0.114. This procedure was repeated across a significant sample size of $10^5$ iterations, creating an extensive dataset. Subsequently, we calculated the mean and standard deviation of these cumulative effective radii to assess the distribution and variability across various directions. Our analysis indicated an average $\Sigma R_{i}= 113$~Mpc, with a standard deviation of 50~Mpc. Furthermore, we evaluated the probabilities of $\Sigma R_{i} \geq $  150~Mpc, 200~Mpc, and 250~Mpc, which were found to be $23.6~\%$, $4.6~\%$, and $0.6~\%$, respectively. Figure~\ref{Monte_Carlo} visually represents the possible cumulative effective radii of voids within spherical regions across specific areas of the sky. It features a histogram that illustrates the probability of encountering varying amounts of voids in these regions,  The red line depicts the probability density function (PDF) smoothly summarizing the distribution trend. This detailed visualization aids in understanding the distribution, variation, and overall density of cosmic voids.
{Additionally, we conducted a Monte Carlo simulation using the same region of the sky to calculate the fraction of the line of sight that traverses a void. Our results showed that the average cumulative effective radius is 12.97 Mpc, with a standard deviation of 19.33 Mpc. Therefore, on average, approximately 2\% of the distance between us and the source is traversed through a void.}

\section{EBL in the Presence of Cosmic Voids} \label{Model}
Our calculations of the inhomogeneous EBL are based on a modified version of the formalism presented in \cite{Finke10}, considering the direct starlight and the re-processing of starlight by dust. For the purpose of a generic study of the effects of cosmic voids along the line of sight to the $\gamma$-ray source (i.e., GRB~221009A), we assume that a spherical cosmic void is located with its center at redshift $z_v$ and radius $R$ between the observer and the $\gamma$-ray source at redshift $z_{s}$. The geometry is illustrated in Figure~\ref{geometry}. 
We calculate the angle- and photon-energy-dependent EBL energy density at each point between the observer and the source by using co-moving coordinates, converting redshifts $z$ to distances $l (z)$. 
However, the voids are vast areas that are almost empty of luminous matter~\citep[see,][]{2009ApJ...699.1252F, Ricciardelli:2014vga}. The matter density at the center of the void can be less by up to 80\% compared to its surrounding areas~\citep[see, e.g., ][]{Ricciardelli:2014vga}. Therefore, for simplicity, we assumed that the contribution to the EBL energy density produced within cosmic voids is negligible. In this work, we set the total emissivity (including both stellar and dust components) within these voids to zero.
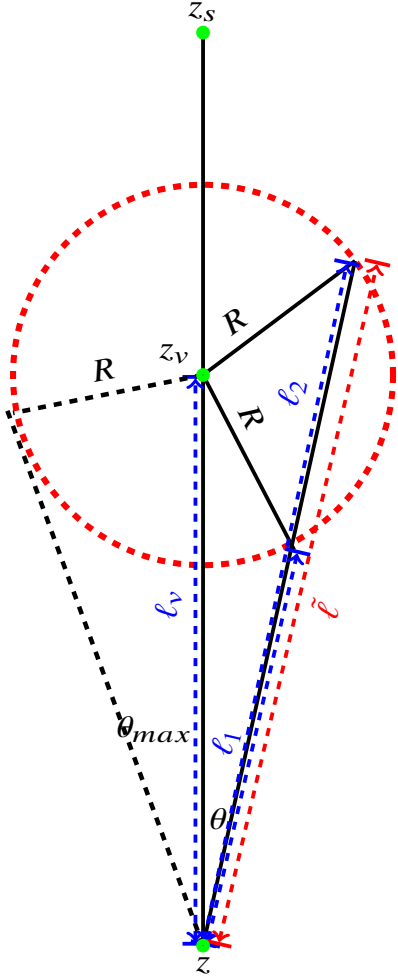
\begin{figure}
\begin{center}
\begin{tikzpicture}[every node/.style={scale=1.5}] 

    \draw [line width=0.8mm, red,dashed] (0,7.5) circle (2.5cm);
   
    \draw [line width=0.5mm] (0,0) -- (0,12);
    \draw[|<->|] [line width=0.5mm, blue, dashed] (-.1,0) -- (-.1,7.5) node[pos=0.6, sloped, above] {$\ell_v$}; 
    \draw [line width=0.5mm] (0,7.5) -- (2,9) node[pos=0.3, sloped, above] {$R$};
    \draw[line width=0.5mm]  (0,7.5) -- (1.2,5.2) node[pos=0.3, sloped, above] {$R$};
    \draw [line width=0.5mm] (0,0) -- (2,9) node[pos=0.15, above] {$\theta~~$};
    \draw[|<->|] [line width=0.5mm, blue, dashed] (0.05,0) -- (1.25,5.2) node[pos=0.5, sloped, above] {$\ell_{1}$};
    \draw[|<->|] [line width=0.5mm, blue, dashed] (0.0,0) -- (1.9,9.0) node[pos=0.8, sloped, above] {$\ell_{2}$};
    \draw[|<->|] [line width=0.5mm, red, dashed] (0.2,0) -- (2.3,9.0) node[pos=0.5, sloped, below] {$\tilde{\ell} $};
    \draw [line width=0.6mm, dashed] (0,0) -- (-2.6,7.1) node[pos=0.35, above] {$ ~ ~ ~ ~ ~ \theta_{max}$};
    \draw [line width=0.6mm, dashed] (0,7.5) -- (-2.5,7.0) node[pos=0.5, sloped, above] {$R$};
    \node[below] at (0,0) {$z$};
    \node[above left] at (0,7.5) {$z_{v}$};
    \draw [line width=0.8mm, green] (0,0) circle (0.5mm);
    \draw [line width=0.8mm, green] (0,7.5) circle (0.5mm);
    \draw [line width=0.8mm, green] (0,12) circle (0.5mm);
    \node[above] at (0,12) {$z_{s}$};    
  \end{tikzpicture}
    \caption{Illustration of an underdense region (void) situated between 
    the observer at redshift $0$
    and the source at redshift $z_{s}$,
   where a $\gamma$-ray photon is emitted from the source and travelling towards the observer.
    The void is assumed to have a radius of $R$, and the redshift at its center is denoted as $z_{v}$. The distances $\ell_{1,2}$ represent the points where the direction of travel of the EBL photon intersects the boundaries of the void. $\ell_v$ is the distance from the photon's position at redshift $z$ to the center of the void, and $\theta_{\rm max}$ represents the maximum angle at which the direction of travel of the EBL photon still intersects the boundary of the void at one point. 
    \label{geometry}
    }
  \end{center}
\end{figure}
From Figure~\ref{geometry} we can find the distances $\ell_{1,2}$ {where the EBL photon travel direction $\Omega = (\theta, \phi)$ crosses the boundaries of the void}, in the case where the $\gamma$-ray photon is not inside the void, as 
\begin{equation}
\ell_{1,2} (\theta) = \ell_v ~\mu ~{ \pm } ~\sqrt{  R^2  - \ell_v^2 \sin^{2} \theta }, 
\end{equation}
where $\mu = \cos\theta$, $\theta$ being the angle between the line of sight and the EBL photon travel direction, and $\ell_v$ is the distance from the position of the photon at redshift $z$ to the center of the void.  
The maximum angle $\theta_{\rm max}$ at which the EBL photon travel
direction still crosses the boundary at one point, is given by:
\begin{equation}\label{max}
\sin\theta_{\rm max} =  \frac{R}{\ell_{v}}.
\end{equation}
and the corresponding distance to the tangential point, $\tilde{\ell} _{\rm max}$, is given by
\begin{equation}
\tilde{\ell}_{\rm max} = \ell_{v} \, \cos\theta_{\rm max}.
\end{equation}
{To calculate the EBL energy density inside the void, there is exactly one point where an EBL photon crosses the boundary of the void, at a distance $\ell_1$ from $z$.  Note that  $z$ represents an arbitrary redshift along the line of sight between the observer and the source, where it equals zero at the observer's location and $z_s$ at the source. To account for the EBL inhomogeneity, we introduce a step function that sets the emissivities inside the void to zero.
For a point at a distance $\tilde{\ell}$ from the location of $z$ (as a function of $\theta$), this step function can be written as}
\begin{equation}
f_{\rm void}^{\rm outside} ~ = ~ 
 \bigg\{
  \begin{tabular}{cc}
   0 &~ ~ if  ~    $\ell_{1} (\theta) ~  < ~  \tilde{\ell} ~ <  ~ \ell_{2} (\theta)$ \\
   1 & ~ ~ otherwise
  \end{tabular}
\label{fvoid_out}
\end{equation}
for points $z$ along the line of sight that are located outside the void, and 
%
\begin{equation}
  f_{\rm void}^{\rm inside} ~ = ~ 
 \bigg\{
  \begin{tabular}{cc}
   0 &~ ~ if  ~    $ \tilde{\ell} ~ <  ~ \ell_{1} (\theta)$ \\
   1 & ~ ~ otherwise
  \end{tabular}
  \ .
\label{fvoid_in}
\end{equation}
We adapt the expression presented in \cite{Finke10} to modify the comoving luminosity density (luminosity per unit comoving volume), also termed as emissivity. This expression allows us to investigate the dependence of the luminosity density on comoving energy $\epsilon$ at a specific redshift $z_e$:
\begin{equation}\label{jstar}
\begin{split}
\epsilon j^{stars}(\epsilon,  z_e) =  m_{e} c^{2} \epsilon^{2} f_{esc} (\epsilon)
  \int_{m_{\rm min}} ^{m_{\rm max}} d m \, \xi(m) \\
  \times  \int_{{z_e}} ^{z_{\rm max}}  d z_{1}\left|\frac{dt_{*}}{d z_{1}}\right| \psi(z_{1}) \dot{N} (\epsilon, m, t_{*}[z_e, z_{1}])\ ; 
\end{split}
\end{equation}
where the initial mass function (IMF) is represented by $\xi(m)$, the comoving star formation rate (SFR) density (i.e., the SFR per unit comoving volume) is denoted as $\psi(z_{1})$, $f_{esc} (\epsilon)$ is the fraction of photons 
that manage to escape a galaxy without being absorbed by interstellar dust, and $\dot{N} (\epsilon, m, t_{*}(z_e, z_{1}))$ is the total number of photons emitted per unit energy per unit time.
 The connection between cosmic time and redshift can be expressed as
\begin{equation}
\left|\frac{dt_{*}}{d z_{1}}\right| = \frac{1}{H_{0} (1+z_{1}) \sqrt{\Omega_m (1+z_1)^3 + \Omega_\Lambda}   }.
\label{dtdz}
\end{equation}
To incorporate the effects of dust and account for the inhomogeneity in the comoving luminosity density, we have made modifications to the expression introduced in \cite{Finke10},
\begin{equation}\label{dust}
\begin{split}
\epsilon j^{dust}(\epsilon,  z_e) =  \frac{15}{\pi^4} \int  d  \epsilon  \left[ \frac{1}{f_{esc}(\epsilon)} - 1 \right] j^{stars}(\epsilon,  z_e) \\
\times \sum_{n=1}^{3} \frac{f_{n}}{\Theta^4 _n} \frac{\epsilon^4}{\exp{\left(\frac{\epsilon}{\Theta_n}\right) - 1 }},
\end{split}
\end{equation}
where $f_{n}$ represents the fraction of the absorbed emissivity reradiated within a specific dust component, $\Theta_n  = k_BT_n / m_e c^2$ denotes the dimensionless temperature of the respective dust component with subscripts n = 1, 2, and 3 corresponding to the warm dust, hot dust, and PAHs, respectively \citep[for more details, see,][]{Finke10}.
The differential photon number density of the EBL at a given redshift $z$ along the travel direction of the $\gamma$-ray photon consists of
the direct contribution from stars throughout cosmic history and the indirect contribution from the absorption and re-emission by dust. The energy density $u_{\rm EBL}$ of the EBL in the comoving frame per unit comoving
volume as a function of $\epsilon$ can be written as: 
\begin{equation}\label{ED}
\epsilon u_{\rm EBL}(\epsilon, z) = 
  \int_{z} ^{z_{\rm max}} d z_e \left|\frac{dt}{d z_e}\right| 
 \frac{\epsilon^{\prime} j(\epsilon^{\prime,}, z_e)}{(1 + z_e)}
\end{equation}
The quantity  $j(\epsilon^{\prime}, \tilde z) $ is the stellar and dust emissivity function defined as $j(\epsilon^{\prime}, \tilde z) =  j^{\rm stars}(\epsilon^{\prime}, \tilde z) + j^{\rm dust}(\epsilon^{\prime}, \tilde z)$
given by Eqs. (\ref{jstar}) and (\ref{dust}). $dt/d \tilde{z}$ is evaluated using {\eref{dtdz}}.
An observer at a redshift $ z > 0 $ would perceive a universe that is $(1+z)^3 $ times smaller than its current size. Photons that we observe today with energy $\epsilon$ would have been seen by them with a proper energy $\epsilon_{p} = (1+z) \epsilon $. Taking into account the cosmic voids described earlier, the energy density expected to be observed can be expressed as:
\begin{equation}\label{EBLinh}
\begin{split}
    \epsilon_{p} u_{{\rm EBL,} p} ({\epsilon_{p}, z, \Omega}) = (1+z)^4
  \int_{z} ^{z_{\rm max}} d \tilde z \left|\frac{dt}{d \tilde z}\right| \\
 \frac{\epsilon^{\prime} j(\epsilon^{\prime}, \tilde z, \Omega)}{(1+\tilde z)} f_{\rm void}(z, \tilde z, \Omega) ,
\end{split}
\end{equation}
{where $f_{\rm void}(z, \tilde z, \Omega)$ is the step function set to be zero inside the 
void 
{{(Eqs. \ref{fvoid_out} and \ref{fvoid_in})}}, and \unboldmath }
$$\epsilon^\prime = \epsilon (1+ \tilde z) = \frac{ (1+ \tilde z) }{ (1+z) } \epsilon_{p}. $$ 

When including the effects of voids, the EBL energy density 
exhibits an angular dependence. Consequently, we calculate the optical depth  for a $\gamma$-ray photon from a source at redshift $z_{s}$ with observed energy $E$, using the angle-dependent EBL photon energy density from Equation (\ref{EBLinh}) as \citep[see, e.g.,][]{Razzaque2009ApJ}:
\begin{equation}\label{opa}
\begin{split}
\tau_{\gamma\gamma} (E,z_s)= c ~~ {\int_{0} ^{z_{s}}} d{z_{1}}  
\left|\frac{dt}{d {z_{1}}}\right|   \int_{0} ^{\infty} d\epsilon_{1} \\
 \oint \ d\Omega \frac{\mu_{\epsilon_{1}}(\epsilon_1, z_1, \Omega)}{\epsilon_{1}} ~ (1-\mu) \sigma_{\gamma \gamma}(s).
\end{split}
\end{equation}
The $\gamma~\gamma$ pair-production cross section $\sigma_{\gamma \gamma}(s)$ can be written as:
\begin{equation*}
\sigma_{\gamma \gamma}(s) = \frac{1}{2} \pi r_{e}^2 (1-\beta_{cm}^2) (3 - \beta_{cm}^4) \ln \left( \frac{1+\beta_{cm}}{1-\beta_{cm}} \right) - 2 \beta_{cm}(2 - \beta_{cm}^2)
\label{sigmagg}
\end{equation*}
where $r_{e}$ is the classical electron radius and $\beta_{cm} = (1-\frac{1}{s})^{1/2}$ is the electron-positron 
velocity in the center-of-momentum (c.m.) frame of the $\gamma\gamma$ interaction, $s = \frac{s_{0}}{2} (1 - \cos \theta)$ 
is the c.m. frame electron/positron energy squared and $s_{0} = \frac{\epsilon E}{m_e^2 c^4} $.
In the case of a homogeneous and isotropic EBL (with which we will compare our results for the inhomogeneous
EBL case), equation (\ref{opa}) can be simplified using the dimensionless function $\bar{\varphi}$ defined by \citet{Gould:1967zzb}:
$$\bar{\varphi}[s_{0}(\epsilon)] = \int_{1} ^{s_{0}(\epsilon)} s \bar{\sigma}(s) ds, $$
where $\bar{\sigma}(s) = \frac{2 \sigma(s)}{\pi r_{e}^{2}}$ and $s_{0}(\epsilon) = E (1+z) \epsilon / m_{e} ^2 c^{4}$,
so that equation (\ref{opa}) can be written as 
\begin{equation}\label{opah}
\begin{split}
\tau_{\gamma\gamma}^{\rm hom} (E,z)= ~  c ~ \pi r_e ^2 \left(\frac{m_{e}^2 c^{4}}{E} \right)^{2} 
{\int_{0} ^{z_{s}}} \frac{d{z_{1}}}{(1+z_{1})^{2}}   \left|\frac{dt}{d {z_{1}}}\right| \\
\times \int_{{\frac{m_{e}^2 c^4}{E (1+z1)}}} ^{\infty} d\epsilon_{1} ~ 
\frac{\mu_{\epsilon_{1}}}{\epsilon_{1}^{3}} ~ \bar{\varphi}[s_{0}(\epsilon)].
\end{split}
\end{equation}
Knowing the optical depth $\tau_{\gamma\gamma}$, we can calculate the attenuation of the intrinsic 
photon flux $F_{\nu}^{int}$ as
\begin{equation}\label{exp} 
F_{\nu}^{\rm obs} = F_{\nu}^{\rm int} e^{-\tau_{\gamma\gamma} (E,z)},
\end{equation}
where $F_{\nu}^{\rm obs}$ is the observed spectrum. In Fig.~\ref{tau_diff_bands} we show the $\gamma\gamma$ opacity for $z=0.151$, the distance to GRB~221009A, and contributions to the opacity by EBL photons of different wavelengths. It is clear that there would be significant attenuation of $\gamma$-ray flux at energies beyond 10 TeV. In Section \ref{res}, we will explore to what extent the presence of voids may decrease the impact on the EBL ${\gamma\gamma}$ absorption in this energy range. 
The primary objective is to examine whether the presence of a cosmic void along the line-of-sight to GRB~221009A may explain the detection of $\gtrsim 10$~TeV photons.

\section{Results and discussion}\label{res}
In this work we hypothesize that the region along our line of sight to GRB~221009A is dominated by voids of various sizes. We assume that the center of the cumulative void is located at a redshift of $z_{v}=0.075$, given that the redshift of GRB~221009A is $z=0.151$. {This assumption allows us to simplify the calculation using a single void where the center of the hypothetical large sphere, is at the midpoint between us and GRB 221009A.}
In \cite{Abdalla17}, we found that the EBL deficit is proportional to the void's size. The impact of several small voids, each with radius $R_{i}$, is roughly equivalent to the effect of a single large void with a radius R equal to the sum of all the small voids radii $\Sigma R_{i}$. Therefore, the single, arbitrarily void radius $R$ used in our calculations in Section \ref{Model} is effectively equivalent to the cumulative radius $\Sigma R_{i}$ used in Section\ref{void_catalog}.
We consider cases with void diameters between 15\% (corresponding to $R = 50$~Mpc) and 80\% (corresponding to $R = 250$~Mpc) of the distance to GRB~221009A. 

In the top panel of Figure~\ref{opaf}, we show the effect of voids of size $50$ Mpc, $100$ Mpc, $150$ Mpc, $200$ Mpc and $250$ Mpc on the $\gamma\gamma$ optical depth compared to the homogeneous case \citep[black solid line,][]{Finke10}. {The bottom panel of Figure~\ref{opaf} shows that the optical depth  $\tau_{\gamma\gamma}$ can be reduced by up to around  10\% and 30\% at 13 TeV for the cases of $R = 110$~Mpc and $R = 250$~Mpc, respectively. Note that the average combined radius that we obtain from our calculations in Section \ref{Monte Carlo} is R $\sim$ 113 Mpc.}
The deficit due to cosmic voids is much higher for TeV photons than for GeV photons. This can be explained by the fact that the TeV photons interact with a broader target photon spectrum, which means that the deficit for very high energy photons is expected to be due to the deficit in the entire EBL spectrum, while the deficit for GeV photons is only due to the deficit in the optical-ultraviolet EBL component. This explanation is clearly depicted in Figure \ref{tau_diff_bands}, where we show $\tau_{\gamma\gamma}$ calculated for EBL photons of different wavelengths. For the same reason, higher energy photons are more sensitive to the values of the Hubble constant when EBL $\gamma \gamma$ absorption is used to constrain cosmology \citep[e.g.,][]{ Dominguez19, Dom2024}.
\begin{figure}
\centering
 \includegraphics[width=\columnwidth, angle=0]{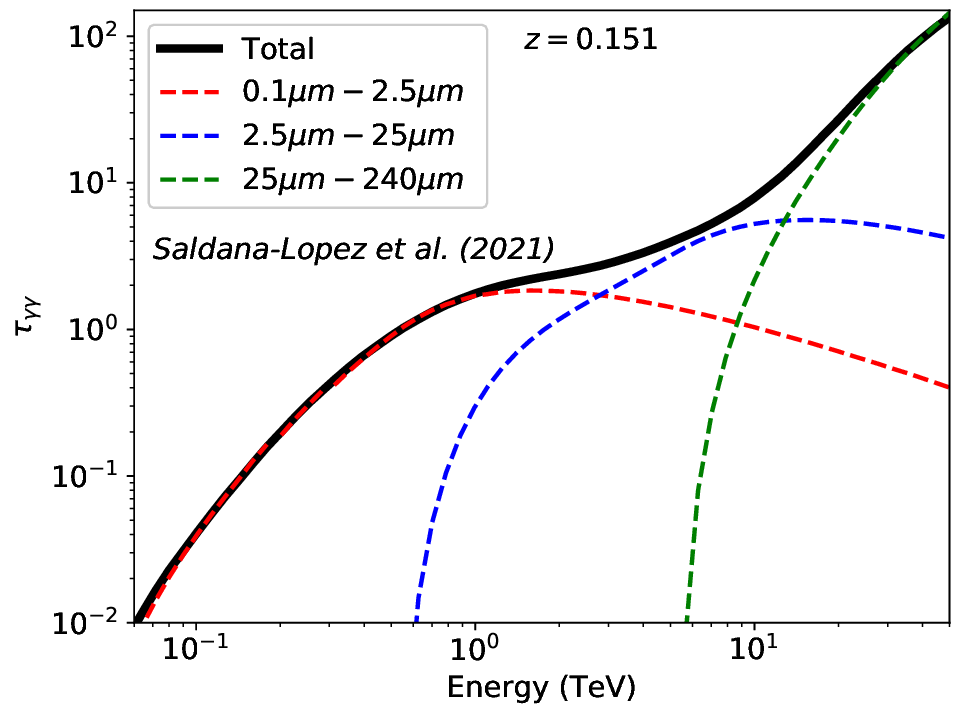}
 \caption{ Optical depth $\tau_{\gamma\gamma}$, for VHE photons traversing from a source located at a redshift $z = 0.151$ attributable to interactions with the EBL, evaluated across various EBL energy bands. For additional details see Figure 10 in \protect\cite{MAGIC_EBL2019}.\label{tau_diff_bands}}
\end{figure}
\begin{figure}
\centering
 \includegraphics[width=\columnwidth, angle=0]{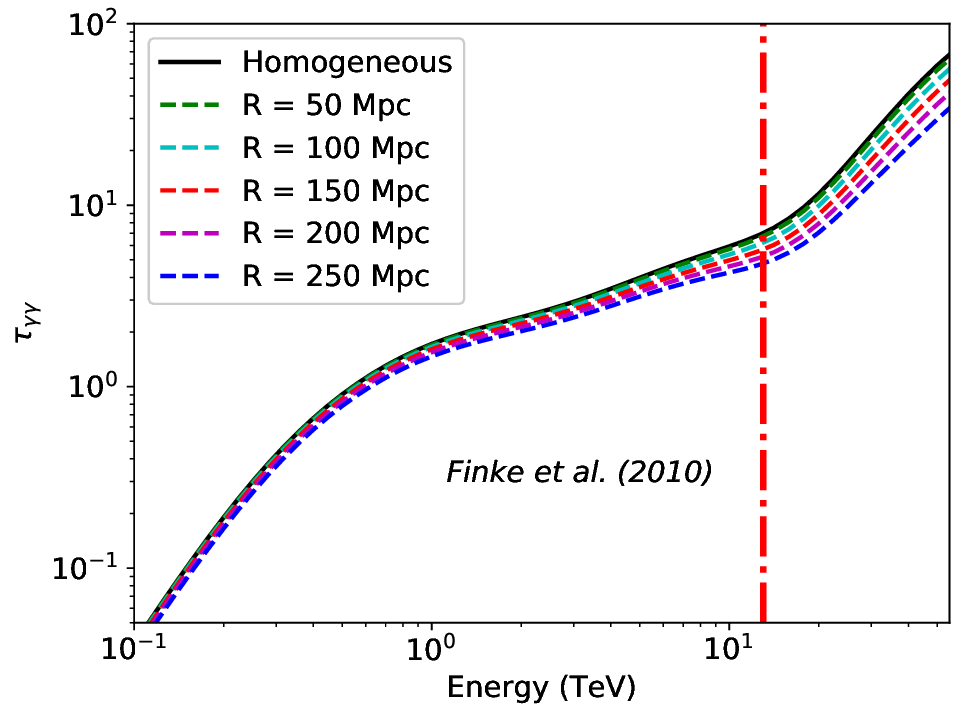}\\
 \includegraphics[width=\columnwidth, angle=0]{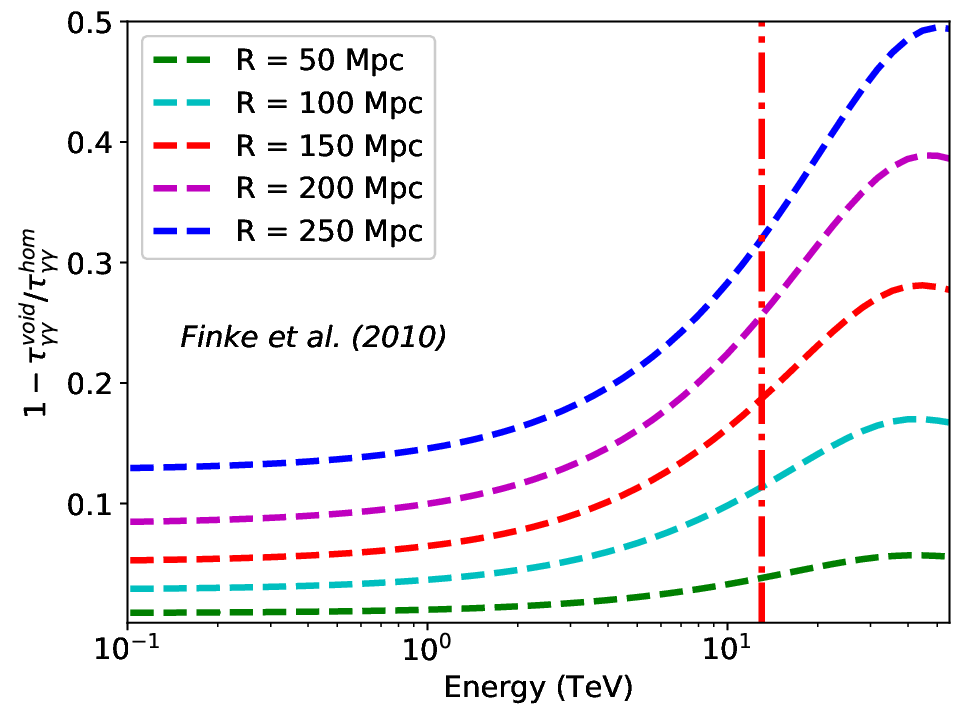}
 \caption{{\bf Top panel:} $\gamma \gamma \to e^+e^-$ optical depth due to the EBL as a function of the $\gamma$-ray energy for a source located at redshift $z = 0.151$. The black solid line indicates the optical depth for a homogeneous EBL distribution. The dashed lines show the opacities in the presence of voids of various sizes, as indicated by the legend. The vertical red dot-dashed line in both panels indicates a photon energy of 13~TeV.
 {\bf Bottom panel:} Relative deficit of the optical depth with and without voids as a function of the $\gamma$-ray energy for the same cases as in the top panel. 
\label{opaf}}
\end{figure} 

\subsection{Cosmic voids using different EBL models}
In Section \ref{Model}, we formulated a methodology for incorporating the concept of EBL inhomogeneity, or the potential existence of cosmic voids, within the framework of the EBL model proposed by \cite{Finke10}. This involved conducting meticulous calculations, assuming that the cumulative emissivity within the void regions could be considered negligible. Incorporating the notion of cosmic voids or under-dense regions within various EBL models is crucial for conducting a thorough analysis and facilitating comparisons with the EBL models featured in \cite{LHAASO2023}. In the top panel of Figure~\ref{opaf} we show the optical depth calculated for the \citet{Finke10} EBL model in the homogeneous case and in the case of cosmic voids of various sizes for a source at $z=0.151$, the redshift of GRB~221009A. The bottom panel shows the relative deficit of optical depth without and with voids.
It is noteworthy that the anticipated effects of cosmic voids are expected to remain consistent across various EBL models. 

To incorporate the effect of cosmic voids into other EBL models we define $\tau_{Finke_{10}}^{hom}$ and $\tau_{Finke_{10}}^{void}$ as the optical depths calculated under two scenarios: one based on the standard homogeneous model and the other considering the presence of cosmic voids, both utilizing \cite{Finke10} EBL model. Similarly, $\tau_{model_x} ^{hom}$ and $\tau_{model_x} ^{void}$ represent analogous optical depths when employing an alternative EBL model denoted as ``$\text{model}_x$'' under the same two scenarios (homogeneous and void-inclusive). Therefore, the optical depth considering voids for any EBL model can be calculated as, 
\begin{equation}\label{void_diffEBL} 
\tau_{model_x} ^{void}= \frac{\tau_{Finke_{10}} ^{void}}{\tau_{Finke_{10}} ^{hom}} \tau_{model_x} ^{hom} \,.
\end{equation}
Based on the hypothesis that the expected optical depth deficit remains consistent regardless of the chosen EBL model, we can introduce EBL inhomogeneity into various EBL models using \eref{void_diffEBL}.
Figure \ref{fig:diffEBL} displays the optical depths for photons originating from GRB 221009A as calculated using the \cite{Dominguez11a}, \cite{Saldana-Lopez23} and \cite{Finke2022} EBL models. 
{The amount by which the cosmic voids lower the optical depth depends on the amount of potential voids that exist along the line-of-sight to the source. Next we study how this change in the optical depth could potentially impact the observed flux for each EBL model, which amount of voids could provide an explanation for the LHAASO data, and how much this could increase the detectability of VHE $\gamma$-rays from these sources.}
\begin{figure}
\centering
\begin{minipage}[b]{.5\linewidth}
\includegraphics[width=\linewidth]{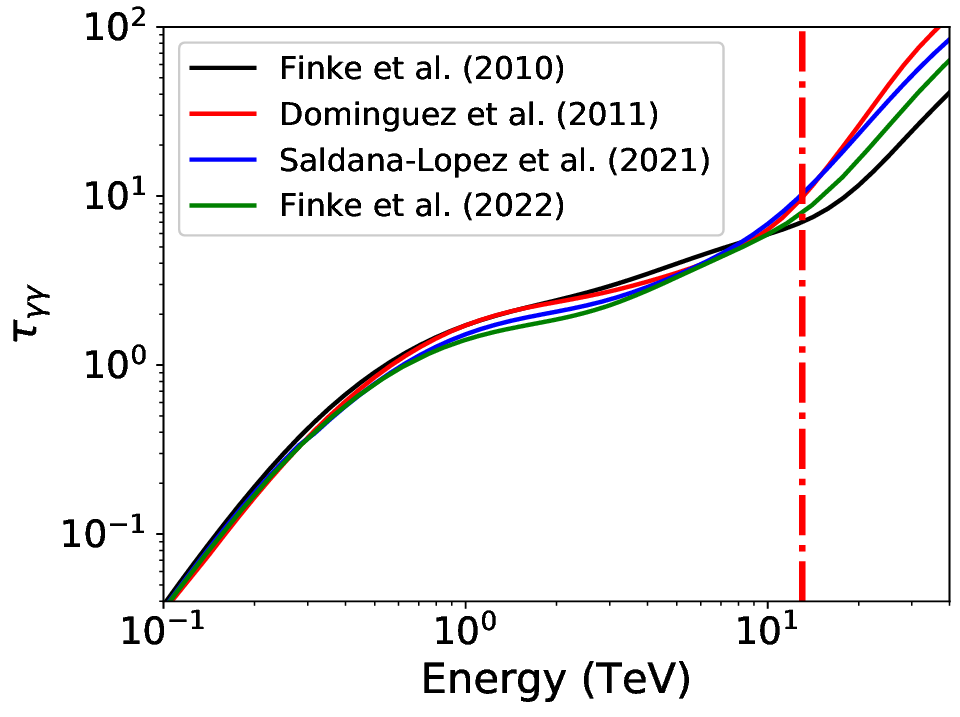}
\end{minipage}%
\begin{minipage}[b]{.5\linewidth}
\includegraphics[width=\linewidth]{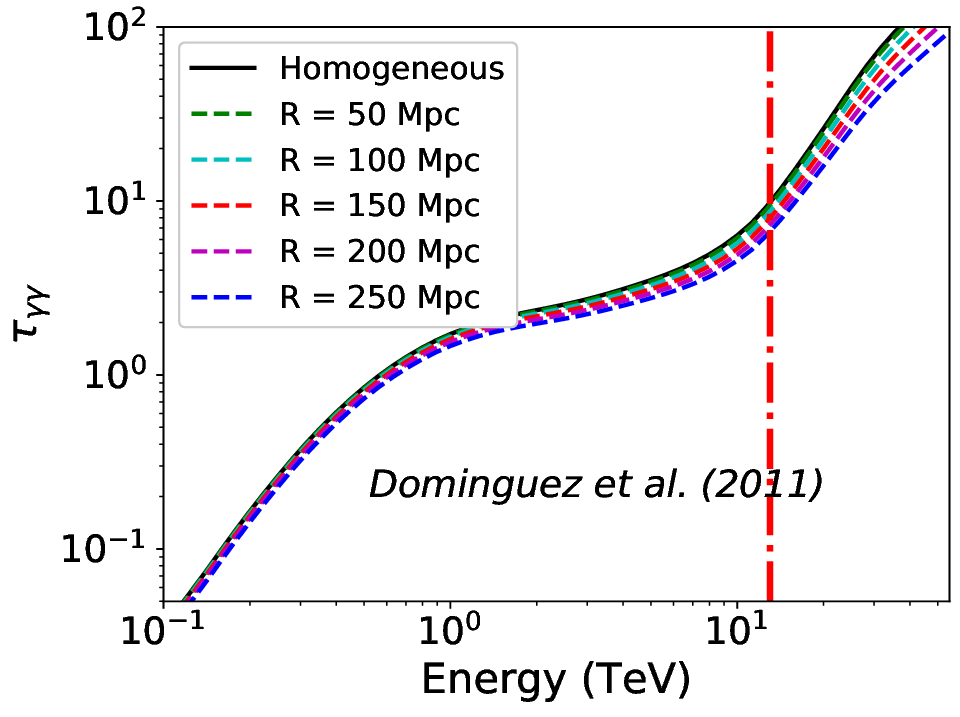}
\end{minipage}
\begin{minipage}[b]{.5\linewidth}
\includegraphics[width=\linewidth]{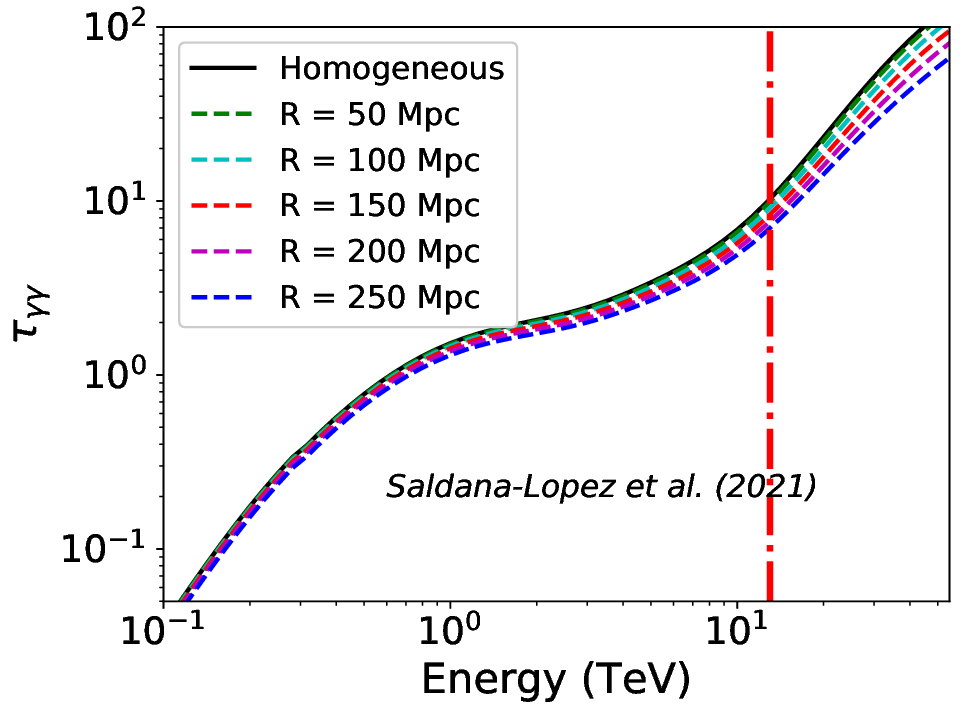}
\end{minipage}%
\begin{minipage}[b]{.5\linewidth}
\includegraphics[width=\linewidth]{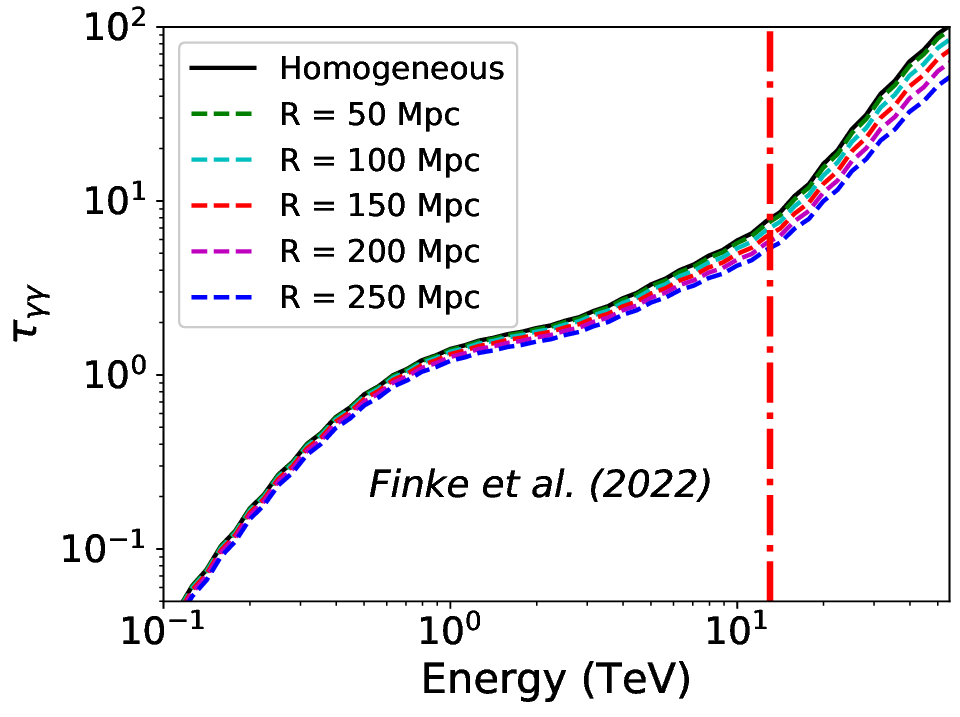}
\end{minipage}
\caption{The optical depth for $\gamma \gamma \to e^+e^-$ interactions in the EBL, as a function of gamma-ray energy, is presented for a source located at redshift $z = 0.151$. In the top-left panel, different lines correspond to various EBL models. We consider both the standard homogeneous case and introduce various void sizes, as illustrated in the corresponding figures. Cosmic voids have been incorporated into the EBL models proposed by \protect\cite{Dominguez11a}, \protect\cite{Saldana-Lopez23}, and \protect\cite{Finke2022}. The vertical red dot-dashed line indicates a photon energy of 13 TeV, which represents the highest energy photon detected from GRB 221009A by LHAASO, as reported in \protect\cite{LHAASO2023}.}
\label{fig:diffEBL}
\end{figure}
\subsection{GRB~221009A: Observed and intrinsic fluxes}
The LHAASO-KM2A detector has observed an unexpected abundance of 140 gamma-ray events with energies exceeding 3~TeV within a time span of 230 to 900 seconds following the initial trigger. A log-parabola function (referred to as LP) and a power-law with an exponential cutoff function (referred to as PLEC) has been employed to describe the intrinsic energy spectrum of these gamma rays, incorporating essential corrections to account for the effects of EBL absorption \cite{LHAASO2023}. 
A detailed analysis of GRB~221009A, accounting for the systematic uncertainties in photon-energy determination, indicates that the maximum energy observed from this source is, depending on the assumed spectrum, approximately 18~TeV (LP) or 13 TeV (PLEC),
as reported by \cite{LHAASO2023}.

In this study, we adopted an arbitrary power-law intrinsic flux and incorporated EBL attenuation, represented by the equation  $dN/dE = N (E/TeV)^{-\alpha} \times e^{- \tau_{\gamma \gamma}}$ utilizing the EBL models from \cite{Finke10}, \cite{Dominguez11a}, \cite{Saldana-Lopez23} and \cite{Finke2022}. Specifically, we employed a fitting procedure to match the attenuated flux from the adopted EBL model to the observed data from \cite{LHAASO2023}, thereby effectively constraining the values of the normalization and index of the power-law function for the intrinsic flux. For each EBL model, we calculated the fitting parameters in the standard homogeneous case first. We then incorporated the cosmic voids of different sizes for each EBL model and identified the best-fit parameters for each case. 
The optimal values of the power-law index $\alpha$ for each scenario are presented in Table \ref{tablex}. It can be noted that $\alpha$ becomes smaller (the spectrum becomes harder) with increasing void size, as expected from a decreasing opacity. However, the value is not significantly different from the spectral index in the homogeneous case ($\boldsymbol{\alpha}_{hom}$).
\begin{table*}
\centering
\caption{The spectral fitting analyses of GRB 221009A, conducted over two time intervals $T_{0} + 230$~s to 300~s and $T_{0} + 300$~s to 900~s, employed a power-law functional form, $dN/dE = N (E/TeV)^{- \alpha}$. The best fitted value for $\alpha$ in the context of homogeneous scenarios, denoted by $\alpha_{hom}$, and in the models accounting for varying cosmic void sizes, represented by $\alpha_{R}$, where $R$ is the void radius. Our analysis revealed that the normalization parameter $N$ for the time interval $T_{0} + 230$~s, the values are as follows: $N = (2.9, ~2.0, ~2.1, ~2.1) \times 10^{-6} $ $TeV^{-1} cm^{-2} s^{-1}$ for \protect\cite{Finke10}, \protect\cite{Dominguez11a}, \protect\cite{Finke2022} and \protect\cite{Saldana-Lopez23} respectively. For the time interval $T_{0} + 300$~s, the values are $N = (4.6,~3.7,~3.3,~3.3) \times 10^{-7} $ $TeV^{-1} cm^{-2} s^{-1}$ for \protect\cite{Finke10}, \protect\cite{Dominguez11a}, \protect\cite{Finke2022} and \protect\cite{Saldana-Lopez23} respectively.}
\setlength{\tabcolsep}{6pt} 
\fontsize{10}{13}\selectfont 
\begin{tabular}{|l|l|l|l|l|l|l|l|}
\hline
\textbf{Time Interval} & \textbf{EBL Models} & \textbf{$\alpha_{hom}$} & \textbf{$\alpha_{50}$} & \textbf{$\alpha_{100}$} & \textbf{$\alpha_{150}$} & \textbf{$\alpha_{200}$} & \textbf{$\alpha_{250}$} \\ \hline
\multirow{4}{*}{230 - 300~s} & \cite{Finke10} & 2.20$\pm$0.07 & 2.19$\pm$0.07 & 2.18$\pm$0.08 & 2.17$\pm$0.09 & 2.16$\pm$0.11 & 2.15$\pm$0.11 \\
& \cite{Dominguez11a} & 2.32$\pm$0.08 & 2.31$\pm$0.07 & 2.31$\pm$0.06 & 2.30$\pm$0.06 & 2.29$\pm$0.05 & 2.28$\pm$0.05 \\
& \cite{Saldana-Lopez23} & 2.40$\pm$0.11 & 2.36$\pm$0.11 & 2.36$\pm$0.09 & 2.35$\pm$0.09 & 2.34$\pm$0.08 & 2.33$\pm$0.07 \\
& \cite{Finke2022} & 2.41$\pm$0.08 & 2.40$\pm$0.07 & 2.39$\pm$0.06 & 2.38$\pm$0.06 & 2.37$\pm$0.06 & 2.36$\pm$0.06 \\ \hline
\multirow{4}{*}{300 - 900~s} & \cite{Finke10} & 2.15$\pm$0.06 & 2.14$\pm$0.07 & 2.13$\pm$0.08 & 2.12$\pm$0.09 & 2.11$\pm$0.09 & 2.09$\pm$0.10 \\
& \cite{Dominguez11a} & 2.30$\pm$0.05 & 2.29$\pm$0.05 & 2.28$\pm$0.05 & 2.27$\pm$0.05 & 2.26$\pm$0.05 & 2.24$\pm$0.05 \\
& \cite{Saldana-Lopez23} & 2.36$\pm$0.06 & 2.35$\pm$0.06 & 2.34$\pm$0.05 & 2.33$\pm$0.05 & 2.32$\pm$0.04 & 2.31$\pm$0.05 \\
& \cite{Finke2022} & 2.38$\pm$0.04 & 2.37$\pm$0.04 & 2.36$\pm$0.04 & 2.35$\pm$0.04 & 2.34$\pm$0.05 & 2.33$\pm$0.05 \\ \hline
\end{tabular}
\label{tablex}
\end{table*}
\begin{figure}
\centering
\begin{minipage}[b]{.5\linewidth}
\centering
\includegraphics[width=\linewidth]{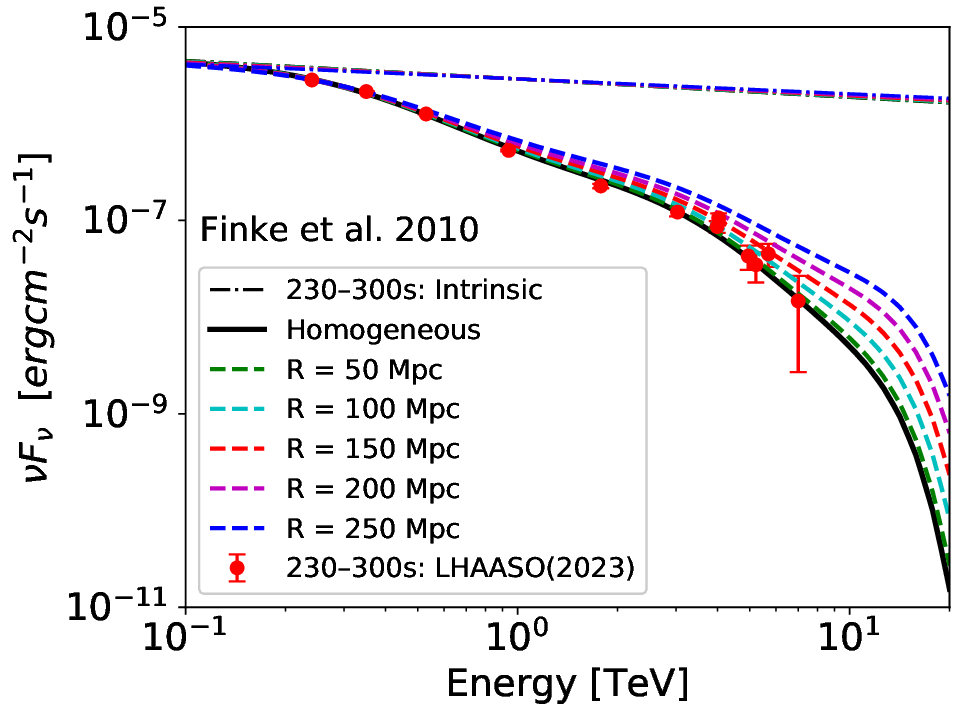} 
\end{minipage}%
\begin{minipage}[b]{.5\linewidth}
\centering
\includegraphics[width=\linewidth]{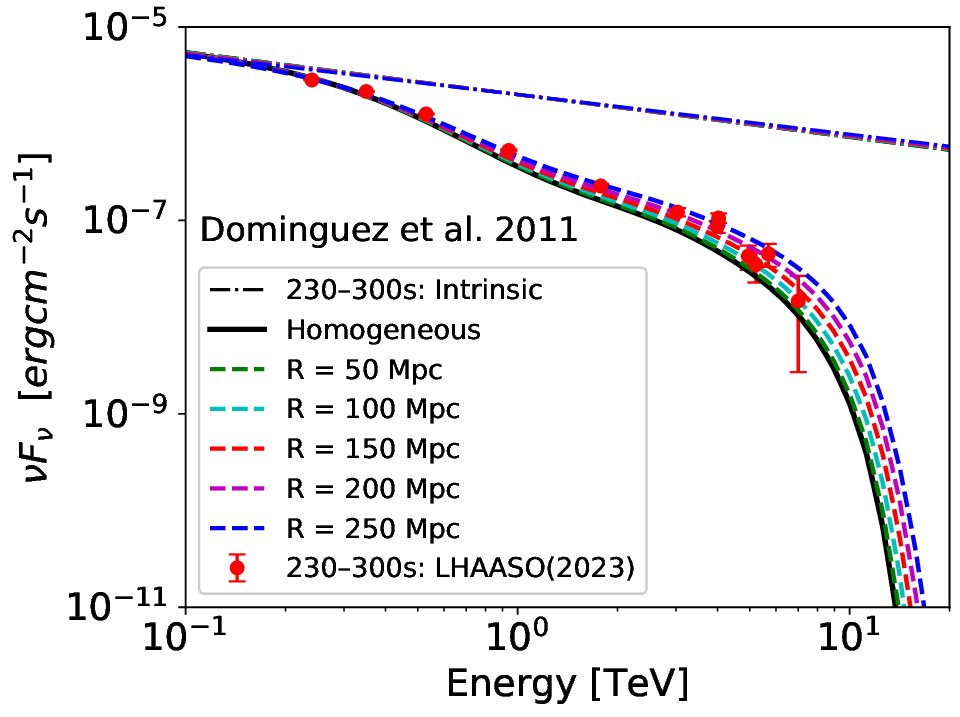} 
\end{minipage}
\begin{minipage}[b]{.5\linewidth}
\centering
\includegraphics[width=\linewidth]{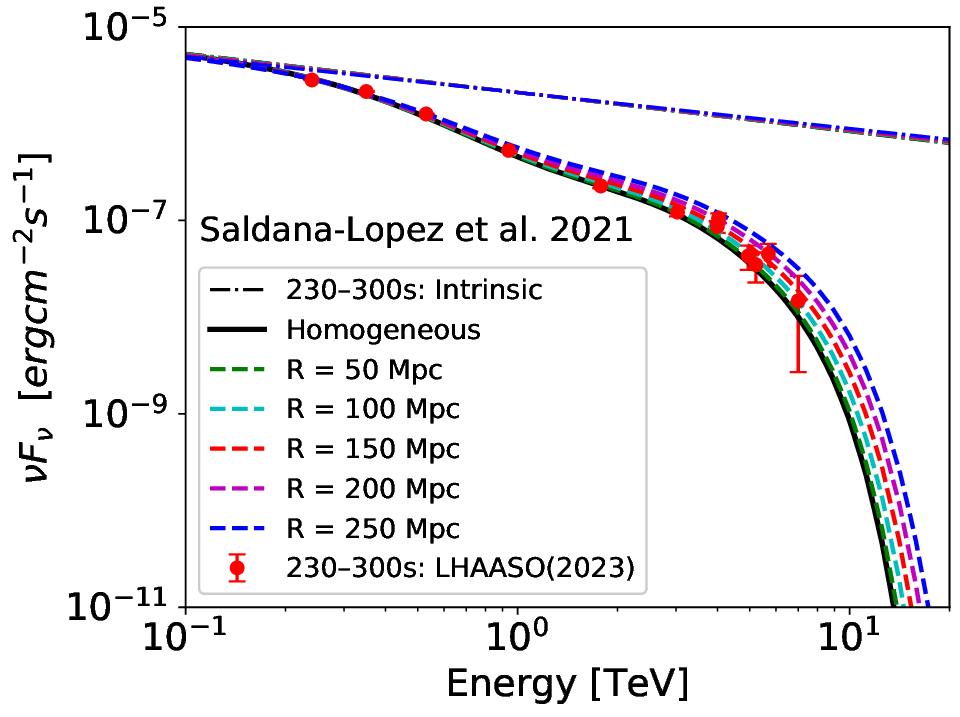} 
\end{minipage}%
\begin{minipage}[b]{.5\linewidth}
\centering
\includegraphics[width=\linewidth]{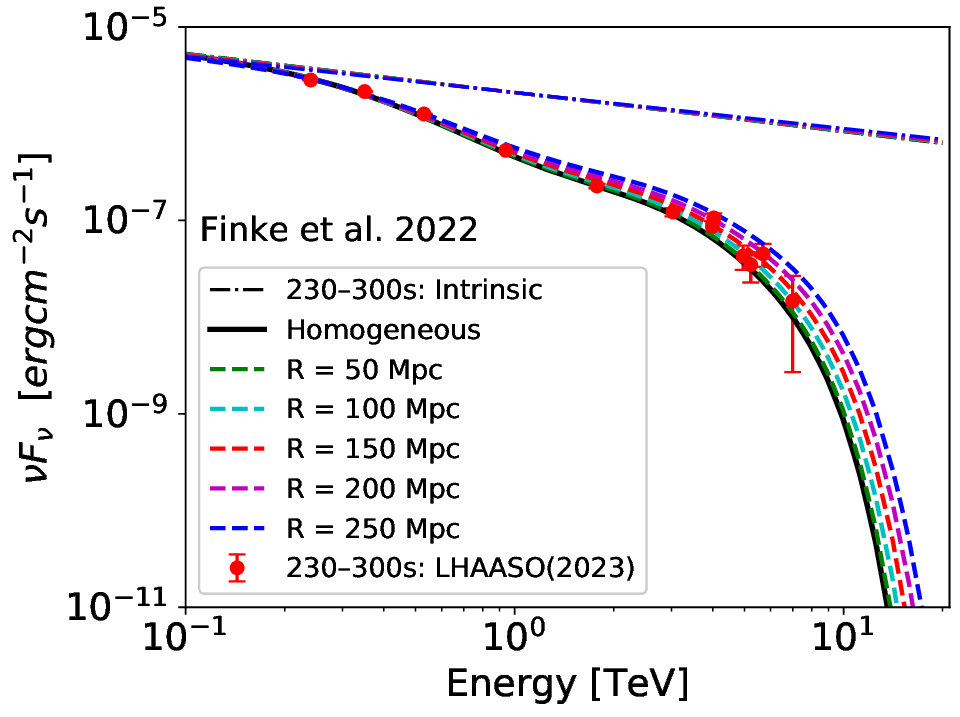} 
\end{minipage}
\caption{The red data points are LHAASO observations for the interval from $T_{0} + 230$~s to $T_{0} + 300$~s, and the intrinsic spectrum obtained through a power-law (PL) model is represented by the solid blue line, employing the EBL models described in \protect\cite{Finke10}, \protect\cite{Dominguez11a}, \protect\cite{Saldana-Lopez23}, and \protect\cite{Finke2022}. The anticipated EBL-induced absorption, as estimated by the indicated EBL models, is represented by the solid black line. Additionally, we explore the influence of cosmic voids by incorporating various void sizes, depicted as dashed lines.}
\label{pl1}
\end{figure}

Figures \ref{pl1} and \ref{pl2} showcase the intrinsic spectra of GRB~221009A, which have been derived using the power-law model based on the fitted parameters detailed in Table \ref{tablex}. These figures also present the anticipated observable flux as estimated by various EBL models. These spectra are shown for two distinct time intervals: $T_{0} + 230$~s {to $T_{0} + 300$~s} (Figure \ref{pl1}) and $T_{0} + 300$~s {to $T_{0} + 900$~s} (Figure \ref{pl2}) as observed by LHAASO~\citep[see,][]{LHAASO2023}. The expected EBL-induced absorption, as estimated by the EBL models in the homogeneous case, is presented as a solid black line. The potential influence of cosmic voids of various sizes (dashed lines) is investigated. The observed flux by LHAASO, especially in the $T_0 + 300$~s to $T_0 + 900$~s (Figure \ref{pl2}) exhibits a hardening of the spectrum above 10~TeV and the fit seems better with some EBL models in the presence of voids. Although, we notice that no single void model can explain the whole spectrum.
\begin{figure}
\centering
\begin{minipage}[b]{.5\linewidth}
\centering
\includegraphics[width=\linewidth]{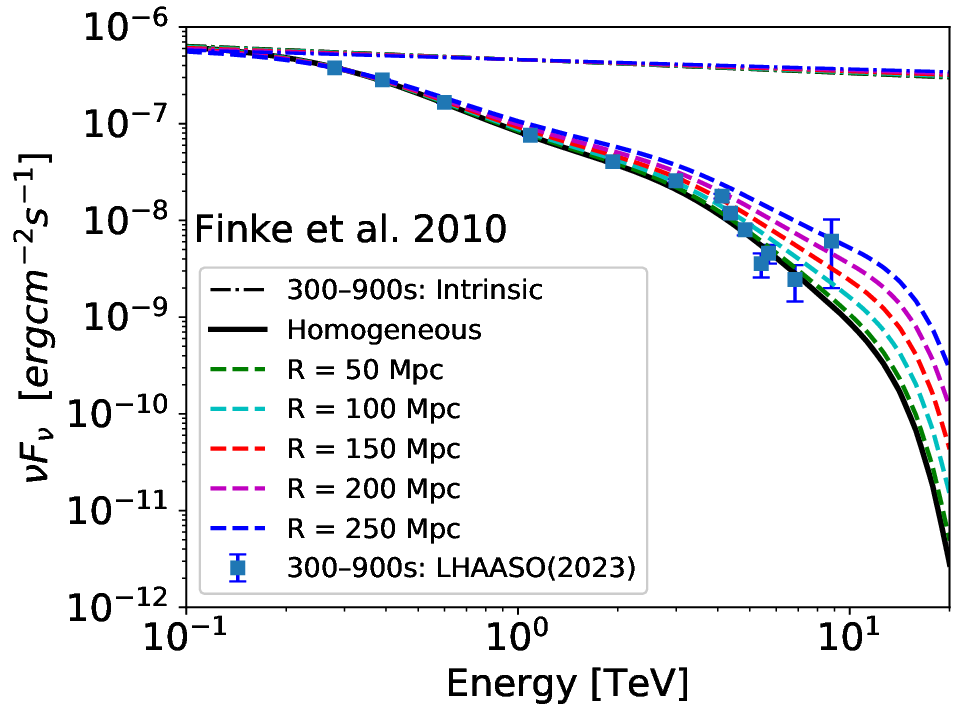}
\end{minipage}%
\begin{minipage}[b]{.5\linewidth}
\centering
\includegraphics[width=\linewidth]{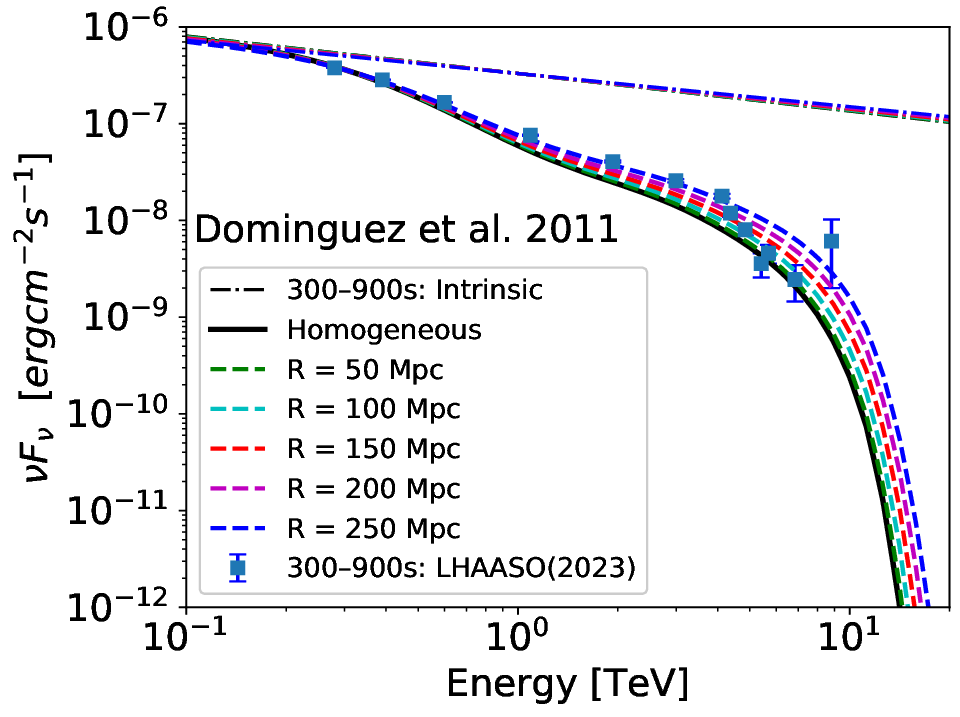}
\end{minipage}
\begin{minipage}[b]{.5\linewidth}
\centering
\includegraphics[width=\linewidth]{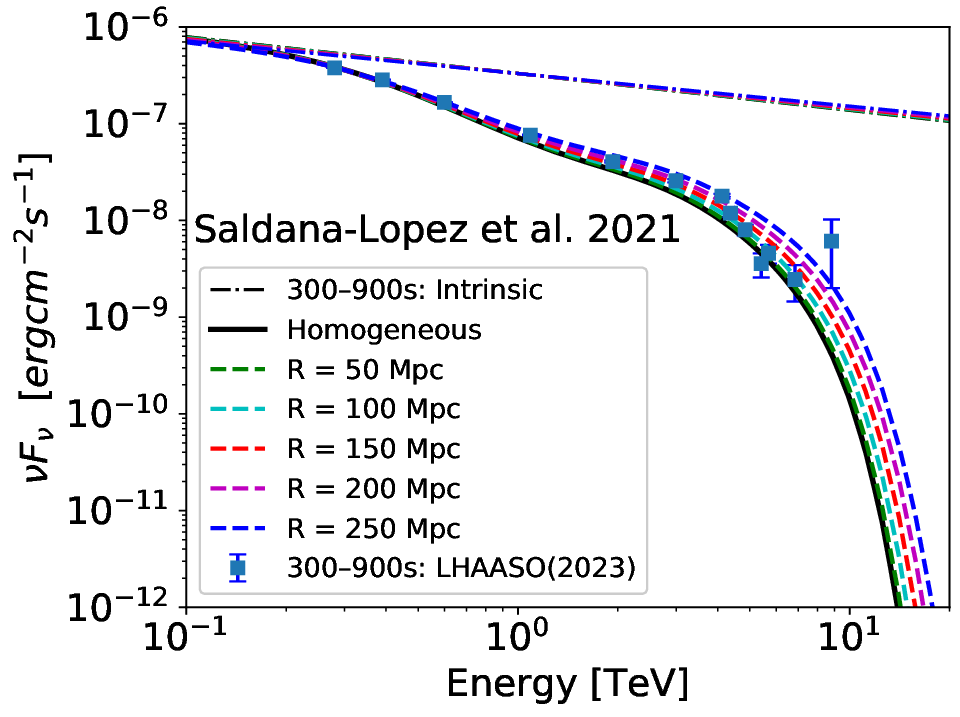}
\end{minipage}%
\begin{minipage}[b]{.5\linewidth}
\centering
\includegraphics[width=\linewidth]{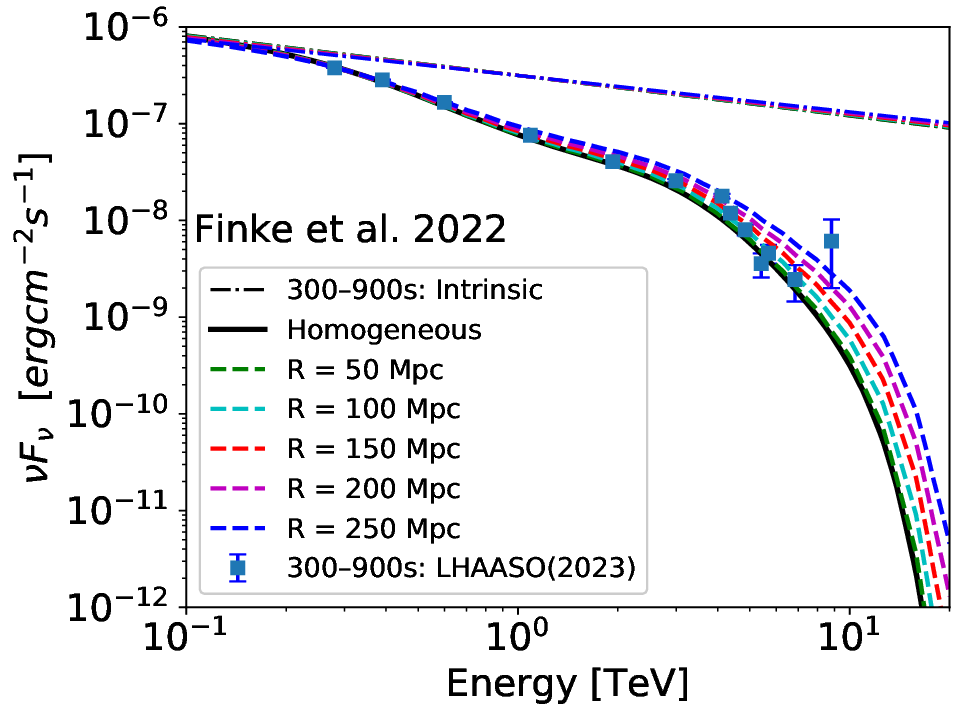}
\end{minipage}
\caption{The blue data points are LHAASO observations for the interval from $T_{0} + 300$~s to $T_{0} + 900$~s, and the intrinsic spectrum obtained through a power-law (PL) model is represented by the solid blue line, employing the EBL models described in \protect\cite{Finke10}, \protect\cite{Dominguez11a}, \protect\cite{Saldana-Lopez23}, and \protect\cite{Finke2022}. The anticipated EBL-induced absorption, as estimated by the indicated EBL models, is represented by the solid black line. Additionally, we explore the influence of cosmic voids by incorporating various void sizes, depicted as dashed lines.}
\label{pl2}
\end{figure}
For voids with accumulated size $R = 250 $~Mpc, the EBL-attenuated flux at 13~TeV is can be greater than a factor of ten than the expected one in the case of a homogeneous EBL. Although illustrative, we note that the LHAASO flux is an optimistic estimate of the source-intrinsic $\gtrsim 10$~TeV flux 
from GRB~221009A, but the reduction of the optical depth is independent of the $\gamma$-ray emission model and intrinsic $\gamma$-ray spectrum. {When combined with our scenario of cosmic voids along the line-of-sight, physical models such as those presented by ~\citet{Das2023A&A...670L..12D, LHAASO23, 2024arXiv240512977Z, Zhang2022arXiv221105754Z} or proton synchrotron emission~\cite{Zhang2022arXiv221105754Z} would require a lower total energy output from GRB 221009A to explain the detection of >10 TeV photons by LHAASO.
}

\subsection{Likelihood Analysis:}
Our goal is to determine how well various EBL models, featuring different void sizes, align with LHAASO's observational data as presented in Figures~\ref{pl1} and \ref{pl2}. We calculated the reduced chi-square statistic for each model with different void sizes, and convert these values to likelihoods for a probabilistic evaluation using the formula $\exp\left(-\frac{\chi_{\text{reduced}}^2}{2}\right)$. \\
\begin{figure}
\centering
 \includegraphics[width=\columnwidth, angle=0]{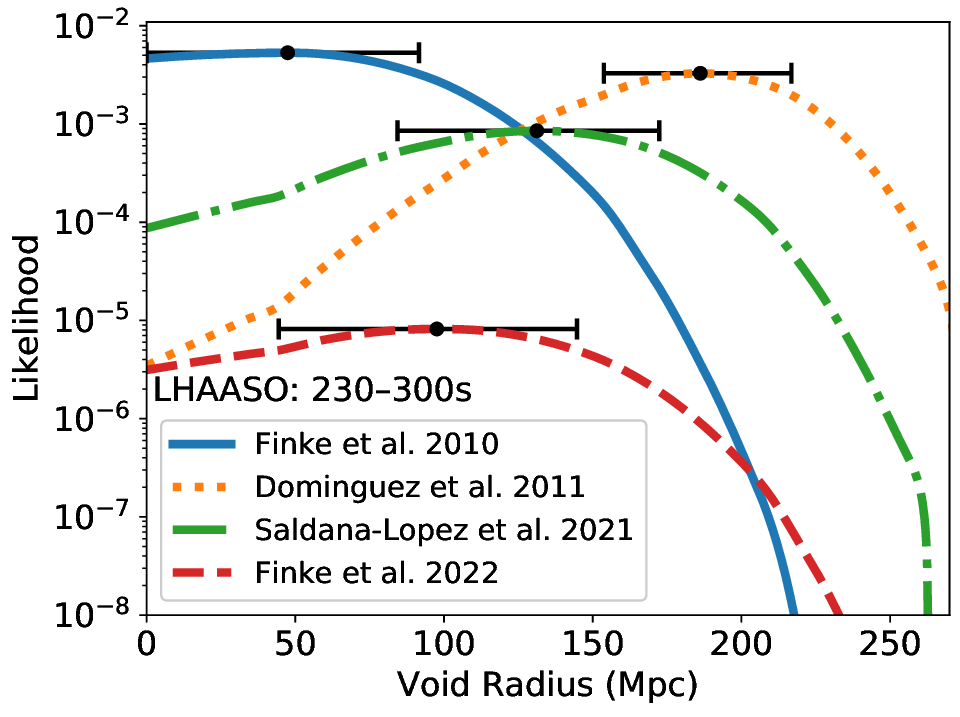}\\
 \includegraphics[width=\columnwidth, angle=0]{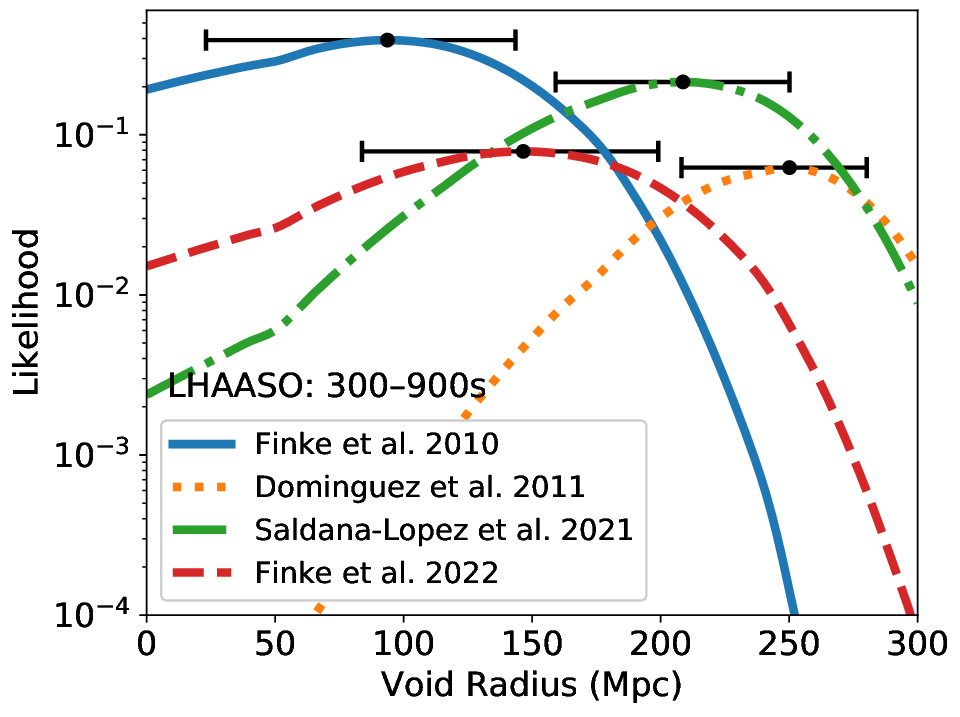}
 \caption{ The likelihood for the dataset from both intervals, $T_{0} + 230$~s {to 300~s} and $T_{0} + 300$~s {to 900~s}, obtained by fitting a Power-Law (PL) model under various EBL models while assuming different quantities of voids in each scenario, is depicted by the solid line. The maximum likelihood value for each model is marked by vertical dashed lines, with colors corresponding to each specific EBL model represented by the solid lines. 
 \label{liklihood} }
\end{figure} 
\begin{table}
\centering
\caption{The cumulative voids radii {$\Sigma R_{i}$} in Mpc necessary for each EBL model get the best fit with LHASSO data.}
\setlength{\tabcolsep}{6pt} 
\fontsize{10}{13}\selectfont 
\begin{tabular}{|l|l|l|l|}
\hline
Time Interval & EBL Models & {$\Sigma R_{i}$} in Mpc  \\ \hline
\multirow{4}{*}{230 - 300~s} & \cite{Finke10} & ~~55±52  \\
                                 & \cite{Dominguez11a} & 215±36 \\
                                 & \cite{Saldana-Lopez23} & 152±51 \\
                                 & \cite{Finke2022} & 113±58 \\ \hline
\multirow{4}{*}{300 - 900~s} & \cite{Finke10} & ~ 94±60 \\
                                 & \cite{Dominguez11a} & 250±36 \\
                                 & \cite{Saldana-Lopez23} & 209±46 \\
                                 & \cite{Finke2022} & 147±58 \\ \hline
\end{tabular}
\label{tab:MLikelihood}
\end{table}
Based on our estimates from the Monte Carlo study (see, \ref{Monte Carlo}), all these various quantities of voids, as required by each model, are possible. However, in certain scenarios, the quantity of voids needed to get the best fit with the LHASSO data appears less likely. Specifically, for \cite{Dominguez11a} and \cite{Saldana-Lopez23} EBL models, the accumulated radii of voids $\Sigma R$ necessary to achieve the best fit can exceed 200~Mpc. The likelihood of such a mount of voids existing for an object at a redshift of $z = 0.151$ is less than 5\%.
\section{Summary and Conclusions}\label{Conclusions}
Cosmic voids remain an active area of research, and their precise sizes remain uncertain from an observational perspective. Determining the minimum size required for a region to be classified as a void, as well as quantifying the matter density compared to the average, poses significant challenges. Estimating the exact reduction in matter density within a spherical region, assumed to be a void, is particularly complex ~\citep[e.g.,][]{Kelly2022}. 
Further in-depth analysis will be crucial in the future to rigorously test our hypothesis and refine our understanding of cosmic voids and their implications\\
In this work, we have explored implications of cosmic voids on the $\gamma$ ray propagation along the line-of-sight and through a region of of reduced EBL opacity, using TeV flux detected by the LHAASO experiment from GRB~221009A. We have computed the optical depth for $\ge 0.1$~TeV photons propagating to the earth from $z=0.151$, the redshift of GRB~221009A. We found that the presence of cosmic voids can reduce the optical depth significantly in the $\gtrsim 1$~TeV range. Correspondingly, assuming the same intrinsic power-law spectrum, the observable $\gamma$-ray flux from GRB~221009A can be greater than a factor of ten for the cases with the larger combined radius compared to the homogeneous EBL models. This can help explain detection of $\gtrsim 10$~TeV photons from GRB~221009A.\\
{We show that the optical depth may be reduced by about 10\% and 30\% at 13 TeV, in the cases of $R = 110$~Mpc and $R = 250$~Mpc, respectively. Furthermore, our findings suggest that the fit of all EBL models is more accurate within the context of possible voids present along our line-of-sight to GRB 221009A}. The quantity of voids needed to achieve optimal fitting falls within the realm of feasibility, as suggested by our Monte Carlo study employing an updated void catalog from VoidFinder SDSS DR7.
Our results, using EBL de-absorption in the presence of void shows a harder intrinsic spectrum than for the homogeneous EBL. Therefore, prospects for future detection of TeV from extragalactic sources such as GRBs or blazars are enhanced, if those take place in the directions where cosmic voids are present.
\section*{Acknowledgements}
 We thank the anonymous referee for a careful reading of the manuscript and helpful suggestions, which have led to significant improvements in the manuscript. The work of H.A. is supported through the María Zambrano researcher fellow funded by the European Union, the AfAS Seed Research Grant, SA-GAMMA ST/S002952/1 and by the University of Johannesburg (UJ) URC grant. S.R.\ was supported by grants from the National Research Foundation (NRF), South Africa, NITheCS, UJ URC. J.D.F.\ was support by NASA through contract S-15633Y and the Office of Naval Research. A.D. is thankful for the support of the Ramón y Cajal program from the Spanish MINECO, Proyecto PID2021-126536OA-I00 funded by MCIN / AEI / 10.13039/501100011033.

\section*{Data Availability}
The codes used to produce the results featured in this article are available upon reasonable request to the corresponding author.


\input{ref.bbl}
\bibliographystyle{mnras}




\bsp	
\label{lastpage}
\end{document}